\documentclass[%
 reprint,
 amsmath,amssymb,
 aps,
]{revtex4-2}

\usepackage{graphicx}
\usepackage{dcolumn}
\usepackage{bm}

\usepackage{calc}
\usepackage{soul}
\usepackage{xcolor}

\begin{document}

\preprint{APS/123-QED}

\title{Ab-Initio Simulation of Field Evaporation}

\author{Jiayuwen Qi}
 \affiliation{Dept.\ of Materials Science and Engineering, The Ohio State University, Columbus, OH, USA}
\author{Christian Oberdorfer}
 \affiliation{Dept.\ of Materials Science and Engineering, The Ohio State University, Columbus, OH, USA}
\author{Emmanuelle A. Marquis}
 \affiliation{Dept.\ of Materials Science and Engineering, University of Michigan, Ann Arbor, MI, USA}
\author{Wolfgang Windl}
 \email{windl.1@osu.edu}
 \affiliation{Dept.\ of Materials Science and Engineering, The Ohio State University, Columbus, OH, USA}

\date{\today}

\begin{abstract}
A new simulation approach of field evaporation is presented. The model combines classical electrostatics with molecular dynamics (MD) simulations. Unlike previous atomic-level simulation approaches, our method does not rely on an evaporation criterion based on thermal activation theory, instead, electric-field-induced forces on atoms are explicitly calculated and added to the interatomic forces. Atoms then simply move according to the laws of classical molecular dynamics and are ``evaporated'' when the external force overcomes interatomic bonding. This approach thus makes no ad-hoc assumptions concerning evaporation fields and criteria, which makes the simulation fully physics-based and {\it ``ab-initio''} apart from the interatomic potential. As proof of principle, we perform simulations to determine material dependent critical voltages which allow assessing the evaporation fields and the corresponding steady-state tip shapes in different metals. {We also extract critical evaporation fields in elemental metals and sublimation energies in a high entropy alloy to have a more direct comparison with tabulated values.} In contrast to previous approaches, we show that our method is able to successfully reproduce the enhanced zone lines observed in experimental field desorption patterns. We also demonstrate the need for careful selection of the interatomic potential by a comparative study for the example of Cu-Ni alloys.
\end{abstract}

\maketitle

\section{\label{sec:intro}Introduction}

Field evaporation is a field-assisted atom removal process in high-electric field nanoscience with applications in liquid ion sources or materials characterization. During field evaporation, surface atoms of a specimen are removed from their original lattice sites and ionized by virtue of a large electric field \cite{miller2014field}. Field evaporation is the central process in atom probe tomography (APT), where a high voltage is applied to a needle-shaped specimen, creating an electric field strong enough to evaporate atoms from the apex. These atoms are then accelerated towards a position-sensitive detector. Their impact time and positions are collected for reconstruction of the data, where the time of flight allows determining the specific atom identity while the combination of arrival time with the $(x,y)$ detector coordinates are used for reconstruction. 

Artifacts are often present and interfere with the fidelity of the reconstruction \cite{ARSLAN2008}, especially in heterogeneous materials and complex microstructures \cite{marquis2011evolution}. However, since the field evaporation process is fully destructive, it is fundamentally impossible to validate the truthfulness of APT reconstruction experimentally. The only viable way to quantify fidelity and uncertainty of reconstruction is via a fully physical computational forward model of field evaporation where every atom is traceable. Consequently, simulation of field evaporation has become an essential tool to detect and interpret artifacts and improve the quality of reconstructed data \cite{vurpillot2013reconstructing}. Physical modelling and simulation of field evaporation has implications beyond the field of atom probe tomography, and can contribute to understanding the electrochemical behavior of surfaces during corrosion \cite{SCULLY2020,Frankel2021} and electrocatalysis \cite{Novaes2021}.

A number of simulation approaches have been developed since the advent of field ion microscopy in the 1950s spanning from continuum to atomistic scales \cite{vurpillot2015modeling}. Initially, simulations were performed to unravel the order of evaporation and the evolution of the surface morphology \cite{moore1962thestructure,suvorov1977computer}. These simulations were based purely on geometry without including field effects. Later continuum models using the level-set method \cite{haley2013level,xu2015simulation,fletcher2019fast} were developed to simulate the evolution of the tip shape in field evaporation \cite{fletcher2020towards}, which provides an alternative way of guiding APT reconstruction by considering the aberration of the tip shape from the hemispherical ideal assumption. These continuum models are computationally efficient and able to model realistic sample geometries. However, because the atomic structure is not included, they are not able to capture the atomic behavior or crystallographic features of the sample such as terraces and facets which are closely related to the trajectories of the evaporated atoms and formation of poles and zone lines in the desorption patterns.

At the other end of the simulation spectrum, first-principles studies of field evaporation within density-functional theory (DFT) can be used to accurately calculate with quantum mechanical approaches desorption energies without and with applied electric field and thus determine the critical evaporation field where the evaporation barrier goes to zero \cite{sanchez2004field,ono2005firstprinciples,peralta2013mapping,karahka2013field,yao2015effects,cui2018onthe,ashton2020abinitio,ohnuma2021firstprinciples}. However, the size of the structural models used in DFT calculations is limited to at most a few hundred atoms and very short time scales, which makes it impossible to examine the full physical process of field evaporation for a realistic sample with realistic electric field. It also restricts these calculations to simplified, slab-like structures.

To fill the gap between the DFT and continuum methods, Vurpillot et al.\ pionnered simulations of a sample with a crystallographic structure and realistic geometry and field distribution \cite{vurpillot1999trajectories}. The atomic structure of an elemental sample tip was modeled by a 3D rigid cubic grid, while the surrounding electrostatic potential was determined within classical electrostatics from Laplace's equation. 
The general procedure had three steps: (1) removal of the surface atom with the highest electric field; (2) recalculation of the electric field distribution for the updated geometry; and (3) calculation of the trajectory of the evaporated atom by integration of Newton's equation of motion. Building on this approach, Geiser et al.\ introduced grids with different sizes, making the model compatible with any cubic structure emitter \cite{geiser2009asystem}. Oberdorfer et al.\ replaced the rigid lattice structure with adaptive unstructured meshes based on Voronoi cells, where the mesh at the tip is determined by the atom positions with increasing coarseness away from the tip \cite{oberdorfer2013afullscale}. Beyond Vurpillot's approach whose use of the Laplace equation restricted modeling materials with infinite permittivity, Oberdorfer solved Poisson's equation to compute the electrostatic potential and electric field. This model now can be applied to any structure with any surface morphology \cite{oberdorfer2011onthefield}. 

Several variations to this general approach have been examined. These include solving the electrostatic field from the Robin equation \cite{rolland2015ameshless}, which has not been widely applied. A kinetic Monte Carlo (KMC) approach to add randomness to the choice of the evaporation event found little effect at experimental temperatures \cite{gruber2011field}. More impactful was the coupling of molecular dynamics (MD) with the finite element method in different simulation approaches \cite{djurabekova2011atomistic, parviainen2015atomistic,oberdorfer2018influence,veske2018dynamic}. There, instead of a ``static'' tip structure, a ``dynamic'' tip structure is thermally equilibrated for several ps after an atom is removed from the tip. Through the addition of surface dynamics, several artefacts could be reproduced and explained. For example, athermal surface migration of solute atoms with much stronger bonds \cite{oberdorfer2018influence}, and rearrangement of the electrostatic field following the field evaporation of atoms \cite{katnagallu2018impact}.

Although the current state of the art in modeling has increased realism by adding atomic structure, realistic field, and atomic motion, one of the most crucial approximations, i.e.\ the convenient but unphysical separation of dynamics from the process of evaporation, has yet to be addressed. To trigger an evaporation event, an ad-hoc criterion is usually introduced, such as comparing the local field at the site of an atom to a tabulated zero-barrier evaporation field (ZBEF) \cite{oberdorfer2013afullscale}. However, Yao et al.\ showed that ZBEF is site-dependent \cite{yao2015effects}. Since ZBEF values are typically related to an activation energy calculated from sublimation energy, workfunction, and ionization energy, Parviainen et al.\ introduced a first site-dependence by determining location-dependent binding energy with classical MD, while keeping ionization energy and work function as global constants \cite{parviainen2015atomistic}. Besides the difficulty of obtaining site-dependent activation energies, evaporation models also ignore the initial detachment velocity of an evaporated atom. It is usually assumed that the initial detachment velocity is zero, thus the atom trajectory always follows the local electric field. Furthermore, it is assumed that an initial velocity component not parallel to the field would  be small and therefore have negligible effects on the detector image except for some minor blurring \cite{vurpillot2018simulation}. In contrast, we show that the component of the electrostatic force not parallel to the interatomic force on the evaporating atom results in noticeable lateral acceleration, and gives rise to detector patterns with enhanced zone lines as we will discuss below.

To address these issues, we present a new field evaporation simulation approach -- ``TAPSim-MD''. It combines the finite element APT simulation package TAPSim \cite{oberdorfer2013afullscale} with the popular molecular dynamics simulation package LAMMPS (Large-scale Atomic/Molecular Massively Parallel Simulator) \cite{plimption1995fast} and replaces criterion-based evaporation with dynamic, force-based evaporation. The simulation is then executed in the form of an MD run where the atoms move under the influence of net forces consisting of the sum of the usual interatomic forces and the electric-field-induced forces. This eliminates the ad-hoc assumptions of an evaporation criterion, the lack of location dependence of the zero-barrier evaporation field, and the zero launch velocity that were part of all previous approaches. Here and in subsequent publications we will show that this approach is more realistic and can provide new understanding of composition dependence of evaporation field, local-environment effects, detailed evaporation mechanisms, their effects on the observed detector images, and artifacts in reconstruction.

\section{\label{sec:method}Methodology}

The electrostatic force calculation follows Oberdorfer's method \cite{oberdorfer2013afullscale}, where the key characteristic was the meshing between Voronoi cells in the vacuum surrounding the tip and the atomic tip structure, where each atom was chosen as the center of a Voronoi cell, while the mesh becomes coarser away from the tip for computational efficiency. Inaccuracies from meshing can be minimized due to the adaptive distribution of Voronoi cells in addition to the emitter cells. After the mesh is established, the electrostatic potential is solved from Poisson's equation on the mesh following the approach described in \cite{oberdorfer2011onthefield}. The electric field is calculated from the electrostatic potential through the numerical differentiation scheme for the irregular grid using the method based on interpolation between the generator points outlined in \cite{oberdorfer2013afullscale}. The polarization charge on each atom is then calculated from Gauss's law (even below the surface in case the material is not a metal and has a finite dielectric constant), and the electrostatic force on each atom is obtained from the product of the charge and the electric field. {In metals that we examine here, the dielectric constant is infinite, which we approximate by a very large value of $10^6$ to keep the numerics stable. As a side note, care needs to be taken for materials with strongly field-dependent dielectric constant such as SrTiO$_3$, where the dielectric constant in the range of evaporation fields is more than one order of magnitude smaller than in the field-free case \cite{Berg95}.}

Oberdorfer's method uniquely determines realistic field-induced force contributions, which when added to the interatomic forces, allows an {\it ``ab-initio''} simulation of field evaporation without specified evaporation criterion. Thus, evaporation is not postulated through a criterion, but observed as a process happening naturally during an MD simulation, where an atom is pulled of the surface by the field-induced force. This process can be detected through vanishing of interatomic forces or by defining a critical velocity for an atom. The sound velocity of the studied material is a convenient value. This thus eliminates the ``selection of surface atoms for desorption'' described in \cite{oberdorfer2013afullscale}. Calculations of trajectories after evaporation and recording of the data then follow again the procedure outlined in \cite{oberdorfer2013afullscale}.

While this approach may sound straightforward, marrying two major codes, especially from the two fundamentally different domains of finite-element modeling and atomic simulation required a large number of handshake steps and streamlining, especially since computational performance is essential to allow for tips that are at least close to realistic dimensions in size. 

{In our MD simulations, we use the microcanonical ensemble ({NVE}), a Langevin thermostat with a damp parameter of 2.50, and the default timestep of 1 fs for metal units. In order to fix the position of the tip base, we set for the atoms in the bottom layer of the virtual tip all velocity components to zero, as well as the $z$-components of their forces. In addition, the ``fix recenter'' command is used to constrain the center-of-mass position of the bottom layer in the $xy$-plane.}

{Regarding the choice of interatomic potentials, we use embedded atom model (EAM) potentials in all but one simulations in this paper to balance the accuracy and speed of computation, since the simulations are computationally rather intensive (see Sec.~\ref{sec:compute}). EAM potentials are known to work very well for close-packed structures and can also work for bcc when the cutoff is long enough. When choosing a potential, we first pick a series of viable candidate potentials based on if their fitting database includes quantities relevant to field evaporation and run test simulations. Then we make decisions based on the assessment of the simulations which should be stable and reasonable. As an example for potential choice, we discuss the process for Cu-Ni alloys in Sec.~\ref{sec:iap}.}

{Based on this, we ended up with potentials by Mishin and co-workers for the fcc-metals Al \cite{alnipotential}, Cu \cite{cupotential}, and Ni \cite{alnipotential} as well as for hcp Co \cite{copotential}. For Au, our potential of choice is by Foiles, Baskes, and Daw \cite{aupotential}. For Ti, which is more challenging becuase of its high-temperature phase transformation from hcp to bcc combined with unique mechanical properties, we chose a very recent potential by Mendelev et al. \cite{tipotential}. While a viable EAM potential was found for bcc W \cite{wpotential}, a comparable potential could not be found for bcc Fe, which then was the only case where we had to resort to a MEAM potential \cite{fepotential}, whose performance due to the included three-body terms is however markedly slower. For the high entropy alloy (HEA), there was only one potential available from Farkas and Caro ~\cite{heapotential}. For validation, we also calculated elemental processes with it as was shown in Fig.~\ref{fig:sim_tsong} and found that they compared favorably to the expectation from Tsong \cite{tsong1978field}.}

\section{\label{sec:validation}Validation of the model}

To validate the new capabilities of the proposed approach, we first demonstrate the electrostatic potential and electric field obtained by Poisson's equation solver. We then compare steady-state tip shapes for different crystal structures of metals and discuss values of critical voltages that for the first time ``TAPSim-MD'' is able to quantitatively evaluate, which allows assessment of evaporation fields across different simulations. We show that the derived average critical evaporation fields correlate strongly with tabulated ZBEFs that are used in classical reconstruction. Alloying effects are investigated in terms of sublimation energies for a high entropy alloy in comparison to its constituent elemental metals. We also show that comparison between the correctly modelled interactomic forces and electric-field-induced-forces explains the origin of enhanced zones lines in field desorption patterns that for the first time successfully reproduce experimental observations.

\subsection{\label{sec:field}Electrostatic potential and electric field}

\begin{figure*}[!htb]
    \centering
    \includegraphics[width=0.9\linewidth]{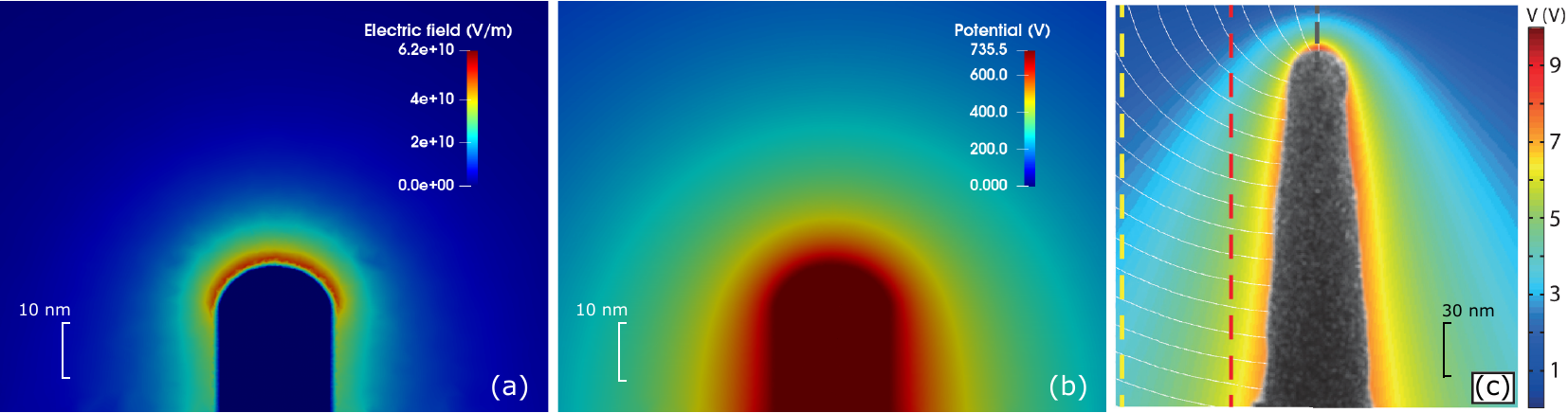}
    \caption{
    {Contour plots of the (a) electric field distribution and (b) electrostatic potential distribution for the center of a 10 nm-radius virtual W tip.}
    {The potential in (b) between two nodes of the mesh is calculated by linear interpolation. The electric field in (a) is computed from the gradient of the potential.}
    (c) is the distribution of the electrostatic potential and electric field around the atom probe needle from electron holography measurements. The white lines represent electric field lines. The colors correspond to equipotential contours. Figure (c) is reprinted from \cite{migunov2015model}.}
    \label{fig:electric_field}
\end{figure*}

The electrostatic potential and electric field distributions are examined in a W tip with the size of 10 nm in radius and 27 nm in height. 
Figure \ref{fig:electric_field} {shows the calculated (a) electric field and (b) electrostatic potential.}
The potential drops rapidly from tip's surface to the vacuum space by about 75\% within 44 nm along the central axis, which results in a strong electric field around the tip. The strongest field with the warmest color is around the top surface of the tip, which is about twice of the field around the side surface of the shaft. 

Migunov et al.\ experimentally determined the electric field and electrostatic potential around a charged tip from a numerical analysis of charge density measurements on the tip surface from electron holography \cite{migunov2015model}. The distribution of the electrostatic potential
and electric field around the atom probe needle is reprinted in Fig.~\ref{fig:electric_field}c. The apex radius of the tip used in the experiment was about 15.6 nm, which is comparable to the virtual tip used in our simulation. The experimental results showed that the potential drops from tip's surface to the vacuum space by about 75\% within 25 nm along the central axis, and the strongest field is found around the top surface of the tip which is about 4 times of the field around the side surface of the shaft. Considering the considerable uncertainties in charge density measurement paired with the strong approximations underlying the numerical analysis, which include perfect cylindrical symmetry of the sample and ellipsoidal shape of the tip, the agreement between our simulations and experiment is good.

\subsection{\label{sec:tip shape}Steady-state tip shape}

In APT experiments, the data collected until a steady-state tip shape is reached are usually discarded. Once the steady-state shape is reached, evaporation of the surface atoms happens typically layer by layer, resulting in a periodically returning tip morphology. Steady-state shapes of virtual tips are also observed in our simulations. Figure 2 includes snapshots of the steady-state shapes of tips of the three most common crystal structures for metals in $\langle 001\rangle$ (or $\langle 0001\rangle$) orientation, {bcc W, fcc Al, and hcp Ti}. The size of the virtual tips is 4 nm in radius. The surface atoms are colored by the magnitude of the electric-field-induced forces (external forces). As expected, the field is concentrated on the edges of the atomic terraces. Therefore, the evaporation of surface atoms always tends to progress from the edges towards the center, which causes faceting. Moreover, the prominent lattice planes revealed by the top view of the tips in Fig.~2 follow the relative prominence rule that states that planes with the largest interplanar spacings are the most prominent \cite{miller2000chapter3}. {For bcc}, the most prominent planes are $\{110\}$, and the second prominent planes are $\{002\}$; {for fcc}, the most prominent planes are $\{111\}$, and the second prominent planes are $\{002\}$; {for hcp}, the most prominent planes are $\{0002\}$, and the second prominent planes are $\{1\overline{1}01\}$ (the most prominent planes in general should be $\{01\overline{1}0\}$, however they are the shaft surface of the cylinder, therefore are not considered here). In Fig.~2, we see indeed $\{110\}$ and $\{002\}$ for W, $\{111\}$ and $\{002\}$ for Al and $\{0002\}$ and $\{1\overline{1}01\}$ for Ti virtual tip as prominent planes respectively. 

\begin{figure*}[!htb]
        \includegraphics[width=0.9\linewidth]{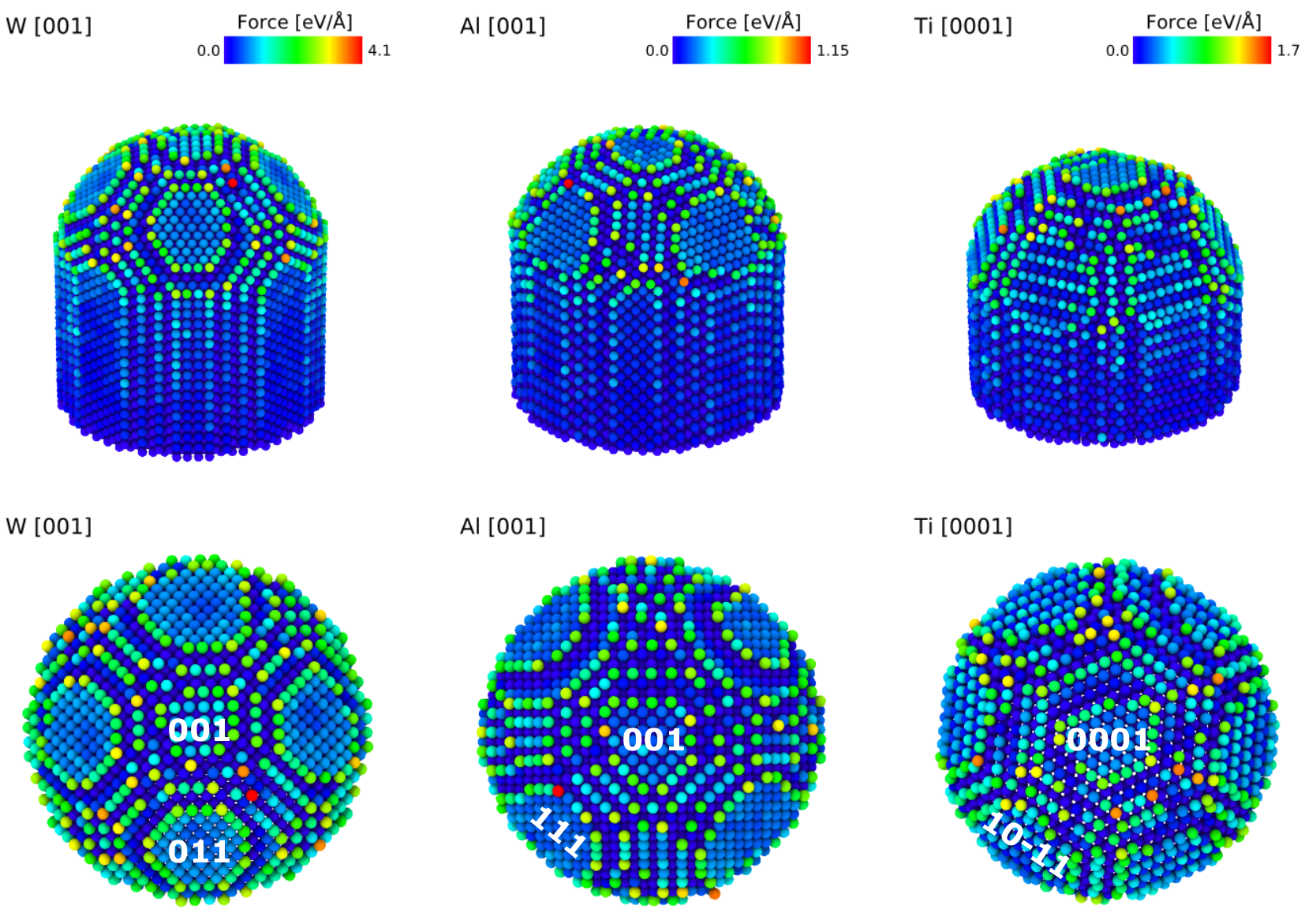}
        \centering
        \caption{Steady-state tip shapes of different metals. Virtual tip's size is 4 nm in radius. Surface atoms are colored by the magnitude of the electric-field-induced forces. The top view of tips are displayed in the second row.}
        \label{fig:elemental_tips}
\end{figure*}

\subsection{\label{sec:voltage}Material-dependent critical voltage}

The applied voltage in previous simulation approaches does not have a quantitative meaning, since codes like TAPSim \cite{oberdorfer2013afullscale} evaporate atoms based on a local field-strength criterion such as the relative strength of the local field vs.\ the tabulated evaporation field of the respective chemical species without considering the local bonding environment. Thus, the voltage is usually set to some reasonable value to create the electric field distribution, irrespective of the material examined.

In our approach, evaporation results from the competition between interatomic forces and field-induced forces, which gives a physical meaning and material dependence to the applied voltage. In order to trigger a single field evaporation event at a time, the applied voltage is controlled to increase first rapidly by 10 V per simulation step then slowly by 0.1 V per simulation step until the first atom evaporates. After that, it is reduced successively by 0.1 V per step until no more evaporation events are detected, and ramped up by 0.1 V per step again until the next evaporation event happens. {Figure \ref{fig:critical_voltages} shows that after some initially high values which we call ``over voltage'' originating from the initially chosen tip shape as discussed in Sec.~\ref{sec:tipshape}, the evaporation voltage oscillates around a constant value which defines the ``critical voltage''.} Its value slightly decreases with decreasing tip's height (keeping the field constant). Since the interatomic forces are determined from carefully fitted interatomic potentials and have realistic values, different materials with different bond strengths will require different voltage values. Also, the same atomic species may require different voltage values when presents in different alloys or compounds where the atomic bonding strength changes and depends on their coordination. Thus, the critical voltage has a physical meaning and can be benchmarked against experimental observations.

\subsubsection{\label{sec:critialvoltage}Critical voltage results}

\begin{figure}[!htb]
    \centering
    \includegraphics{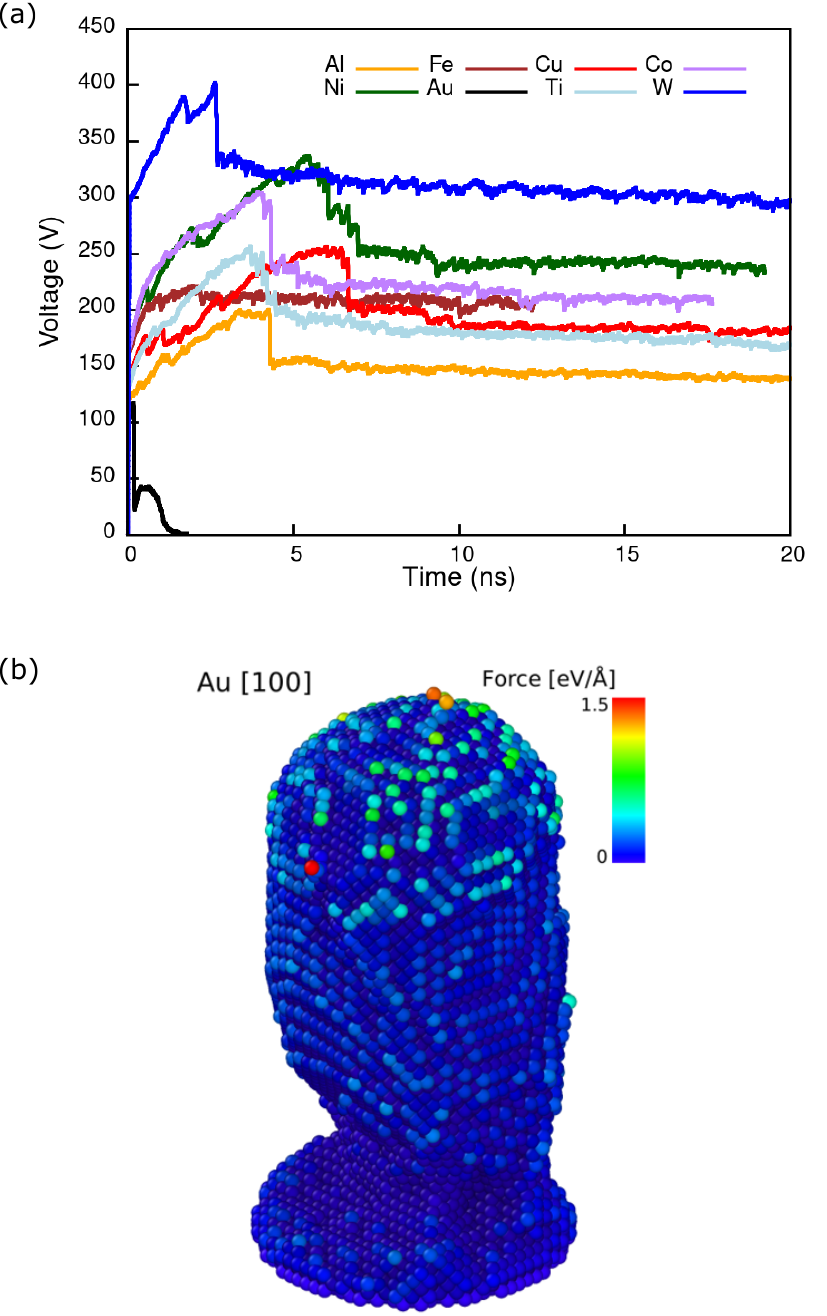}
    \caption{(a) Critical evaporation voltages of different metals. The hemispherical tip size is 4 nm in radius and 12 nm in total height. (b) An Au tip deforms during the simulation which leads to fracture. Surface atoms are colored by the magnitude of the electric-field-induced forces.}
    \label{fig:critical_voltages}
\end{figure}

{The results in Fig.~\ref{fig:critical_voltages} are from  virtual tips in the shape of a hemisphere on top of a cylinder to mimic the idealized shape assumed for classical reconstruction.}
The size of the virtual tips is 4 nm in radius and 12 nm in total height. The $\langle 001\rangle$ direction of the tip ($\langle 0001\rangle$ direction for HCP structure) is oriented along the $z$ axis. Simulated elemental metals as shown in Fig.~\ref{fig:critical_voltages}a include the {fcc metals Al, Cu, Ni, and Au; hcp metals Co  and Ti; and bcc metals W and Fe.}  
 
{The sequence of critical voltages from high to low is W ($\sim295.0$~V), Ni ($\sim241.5$~V), Co ($\sim215.3$ V), Fe ($\sim208.9$ V), Cu ($\sim184.4$ V), Ti ($\sim170.7$ V), and Al ($\sim137.6$ V), which are obtained respectively as the averages of their voltages within the third quarter of the corresponding time ranges. The sequence of the tabulated ZBEF in Tsong's paper \cite{tsong1978field} from high to low is W (5.7 V/{\AA}), Co (3.6 V/{\AA}), Ni (3.5 V/{\AA}), Fe (3.5 V/{\AA}), Cu (3.0 V/{\AA}), Ti (2.5 V/{\AA}), Al (1.9 V/{\AA}).} Despite the reversed sequence of Ni and Co, the simulation results are otherwise in qualitative agreement with the theoretical predictions, which is a first indication that the use of interatomic forces and field-induced forces is able to capture the basic features of field evaporation. The simulation of Au went to failure during the simulation as shown in Fig.~\ref{fig:critical_voltages}b, due to Au's high ductility and peculiar deformation mechanism, which also makes it very difficult to stabilize the pure gold specimen in experiments \cite{ast1968thefield}. While it is not easy to convert the voltage value from the simulation to a corresponding ZBEF due to the geometry dependence of electric fields and the very different, much smaller geometry of the virtual setup in comparison to experiments, we will show in {Sec.~\ref{sec:sublimation energy}} that an approximate virtual evaporation field can be determined from the local forces and benchmarked against tabulated values in a more direct way.

\subsubsection{\label{sec:tipshape}Effect of tip shape on critical voltage}

\begin{figure*}[!htb]
        \includegraphics[width=0.9\linewidth]{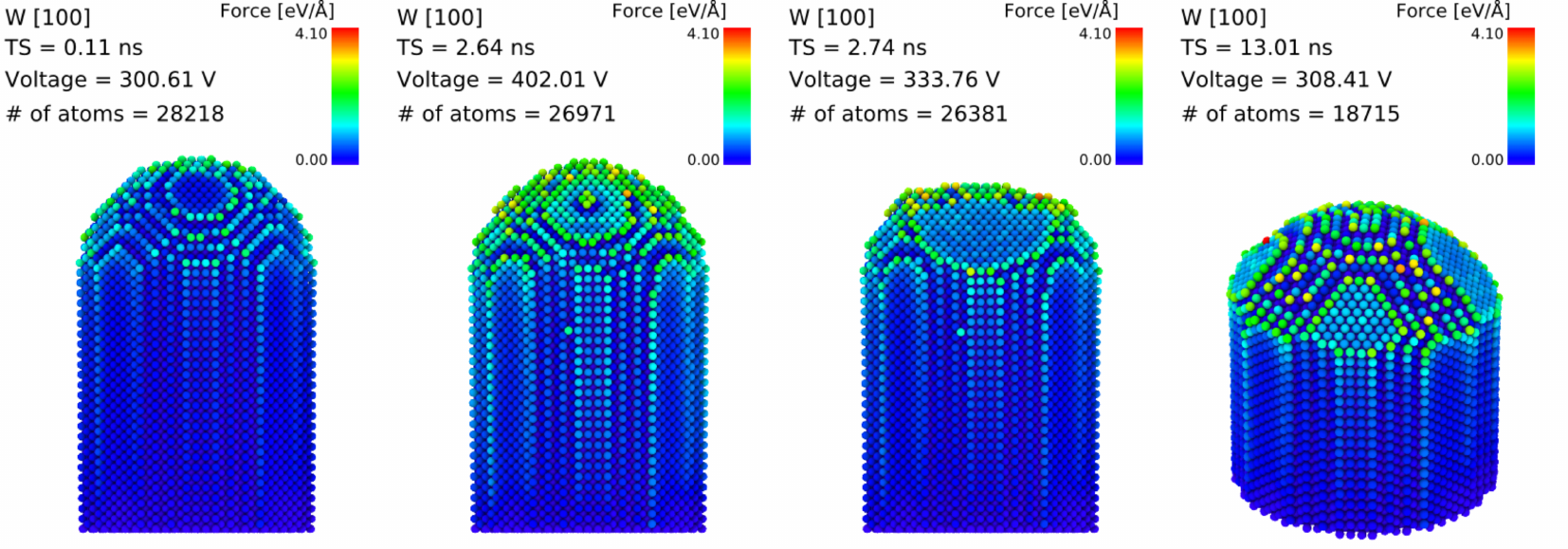}
        \centering
        \caption{Evolution of a 4 nm-radius $\langle 001\rangle$ hemispherical W tip. Surface atoms are colored by the magnitude of the electric-field-induced forces.}
        \label{fig:w_hemisphere_tips}
\end{figure*}

As can be seen in Fig.~\ref{fig:critical_voltages}a, there is a considerable over-voltage stage during the initial shape evolution that lasts a few ns, whose value is about 20\% to 25\% higher than the obtained critical voltage. We find that this over-voltage is closely related to the initial hemispherical tip shape and causes inefficiency in the simulation. Figure \ref{fig:w_hemisphere_tips} is an example of a 4 nm-radius $\langle 100\rangle$ oriented W tip starting with a hemispherical shape. The figure includes snapshots of the tip at 0.11 ns, 2.64 ns, 2.74 ns and 13.01 ns. Since the tip surface is hemispherical, there is no pronounced concentration of electric field on the surface in the beginning, as indicated by the homogeneous coloring of the surface atoms observed up to 2.64 ns. In order to create higher local electric fields and trigger evaporation events, the voltage needs to increase. In this example, due to the stable hemispherical shape, although the voltage goes up considerably by about 34\%, only about 1250 atoms are evaporated during the first 2.53 ns. However, during the following 0.1 ns, about 600 atoms are evaporated very quickly, which results in a very large $\{002\}$ and some smaller $\{110\}$ facets. {The $\{002\}$ terrace has a large edge} where the electric field can concentrate, and the applied voltage drops promptly by about 17\% within 0.1 ns approaching the critical voltage. The tip then evolves to a semi-stable shape as shown at 13.01 ns. Since this avalanche of atoms cannot be analyzed in a meaningful way, the initial computation wastes considerable computer time. The effect is exacerbated as the number of atoms ripped off in the avalanche is too large and results in a too-flat tip surface with a much smaller curvature compared to the steady-state shape, therefore more steps are required before the final tip shape is reached. In this example, the tip is not able to evolve to its steady-state shape shown in Fig.~\ref{fig:elemental_tips} before the whole tip is evaporated. Therefore, a starting hemispherical shape, while perhaps intuitive since close to the faceted steady-state shape, is highly inefficient.

\begin{figure}[!htb]
        \includegraphics{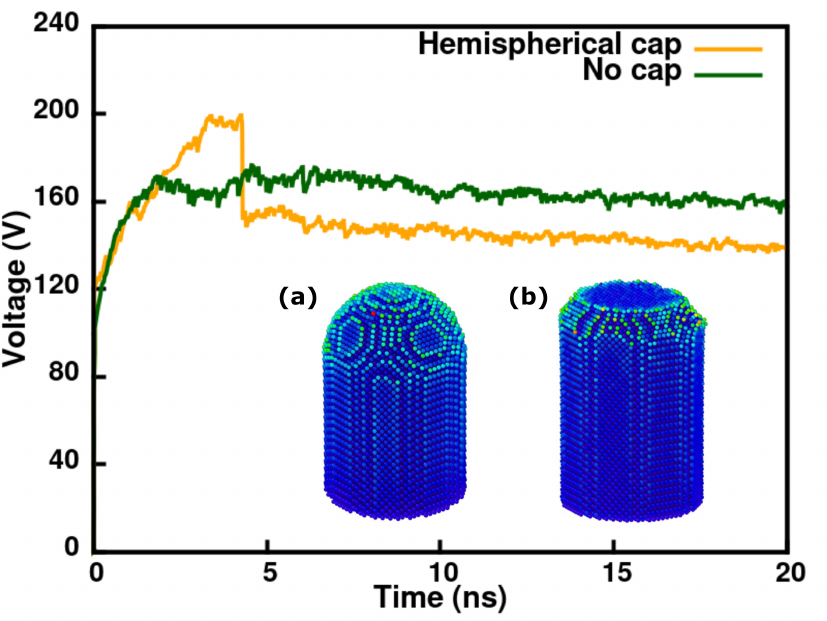}
        \centering
        \caption{Applied voltage vs.\ time in hemispherical and cylindrical Al virtual tips. The tip structure of each case is (a) $\langle 001\rangle$ oriented hemispherical tip and (b) $\langle 001\rangle$ oriented cylindrical tip. The size of both tips is 4 nm in radius. Surface atoms are colored by the magnitude of the electric-field-induced forces.}
        \label{fig:al_voltage_tips}
\end{figure}

In order to overcome this inefficiency, we found a cylinder tip with a flat end cap to be a much better initial structure. We demonstrate this for a $\langle 001\rangle$ oriented Al tip with 4 nm radius. We compare the evolution of evaporation voltage between a cylindrical and hemispherical tip in Fig.~\ref{fig:al_voltage_tips}. In contrast to the voltage applied to the hemispherical tip, the voltage applied to the cylindrical tip quickly reaches its critical voltage within about 2 ns, thanks to the concentration of electric field on the rim of the cylinder. Starting from the edge, the tip gradually evolves to its steady-state shape. There is no ``over-voltage'' stage or avalanche-evaporation of atoms in this process and the steady-state shape is  reached within short simulation time. Therefore we use cylinders as start shapes of virtual tips in all other simulations. Different critical voltages obtained in the two cases result from different tip heights when the voltages reach the respective plateaus. In addition, since a cylinder is distinctly different from the final steady-state shape, the consistent evolution of the tip shape to its steady state in independent simulations further validates the present simulation method.

\subsection{\label{sec:sublimation energy}Critical evaporation field and average sublimation energy}

We derive critical evaporation fields of different elemental metals from simulation results to have a direct comparison to tabulated ZBEF data. 
{To examine alloying effects in an extreme case, simulations are conducted in an fcc Fe$_{18.8}$Ni$_{21.9}$Cr$_{18.9}$Co$_{21.8}$Cu$_{18.6}$ high entropy alloy, which is later referred to as the FeNiCrCoCu HEA or HEA for simplicity. The tip was cut from the periodic replica of a 32-atom special quasi-random structure (SQS) cubic cell with 6 Fe, 6 Cr, 6 Cu, 7 Ni and 7 Co.} 
We compare the obtained sublimation energies to the tabulated data, as well as the sublimation energies and critical voltages for the different elements in their elemental form and in the HEA, where a ``mixed'' critical voltage of the HEA is found.

\begin{figure}[!htb]
    \centering
    \includegraphics{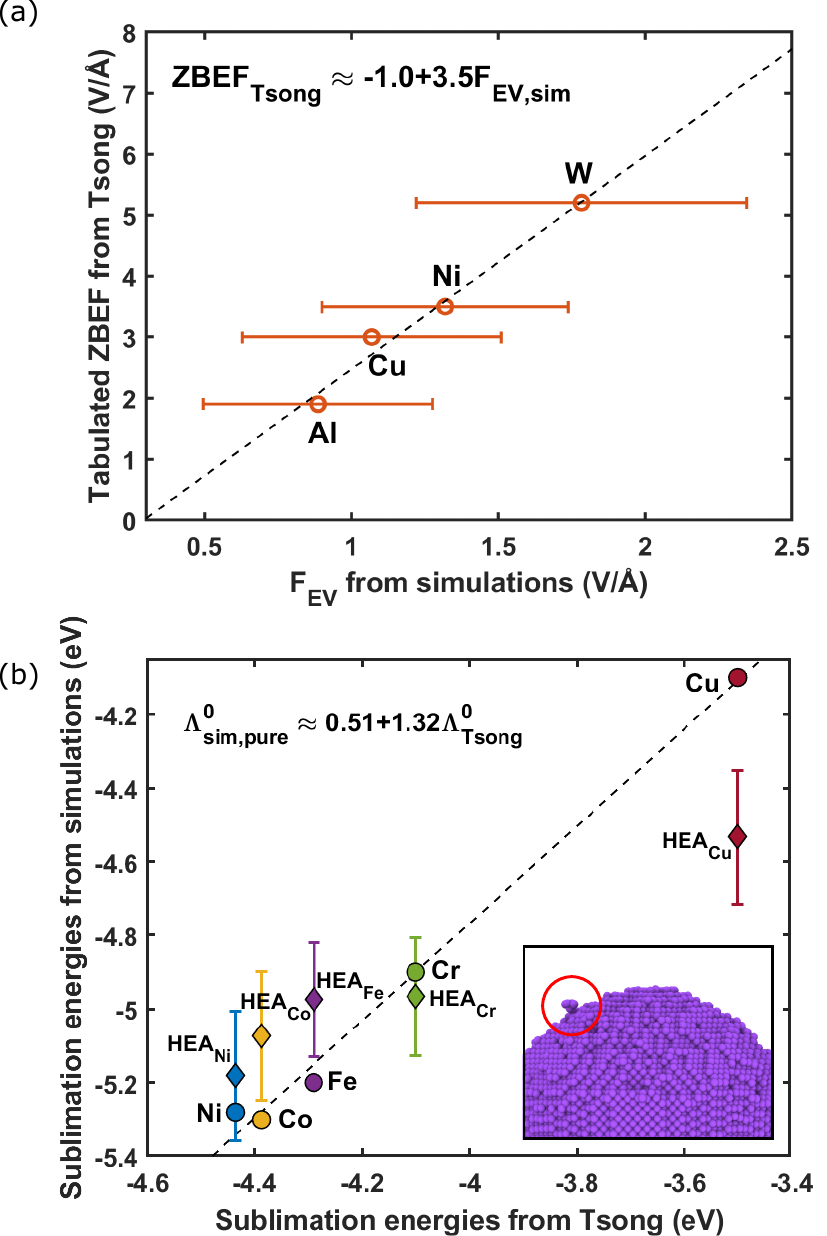}
    \caption{(a) Average critical evaporation fields of different elemental metals from simulations of virtual tips with 4 nm in radius vs.\ tabulated ZBEF values from Tsong \cite{tsong1978field}. The error bars represent {the standard deviation} of the observed evaporation fields. (b) Averaged sublimation energies of Co, Ni, Fe, Cu, and Cr metals in elemental form (circles) and as components of an HEA (diamonds) vs.\ tabulated sublimation energies from Tsong \cite{tsong1978field}. Symbols in the same color represent the same species. 
    {The error bars represent the standard deviation of sublimation energies of components in the HEA.}
    The virtual tips were 5 nm in radius and all simulations used the same interatomic potential. {The inset is the tip surface during evaporation. An evaporating cluster is indicated by a red circle.}}
    \label{fig:sim_tsong}
\end{figure}

Although the local electric field is site-dependent for surface atoms, in order to compare to the tabulated ZBEF in Tsong's paper \cite{tsong1978field}, the critical evaporation field is derived as an averaged value over all evaporation events. The local electric field of an evaporated atom is chosen as its critical evaporation field at the time when its interatomic force goes to zero. As shown in Fig.~\ref{fig:sim_tsong}a, typical tabulated ZBEF values \cite{tsong1978field} are about 3.5 times higher than the critical electric field obtained in our simulations. Since the aberration is systematic, it demonstrates that while the reduced scales of the MD setup cause some scaling in potential and fields, the process of evaporation is described with high fidelity with linear dependence between simulated and calculated values.

A similar analysis has also been done for zero-field sublimation energies of surface atoms {of the FeNiCrCoCu HEA cylinder tip.} 
{Simulations were performed for the HEA tip} of 5 nm in radius as well as for elemental metals of each alloy component using the same interatomic potential. Sublimation energies are computed for $\langle 001\rangle$ surface atoms. In Fig.~\ref{fig:sim_tsong}b, circles are results obtained for elemental metals of HEA components, while diamonds are obtained in the HEA alloy. Symbols of the same element are in the same color. Sublimation energies of elemental metals are very close to each other in the case of Co, Fe, and Ni, whereas Cu bonds are much weaker than the others. 
{Compared to the homogeneous bonding environment in the elemental metals, there is a distribution of sublimation energies for each of the components in the HEA as shown in Fig.~\ref{fig:sim_tsong}b, which indicates a mixing effect caused by different bonding contributions from different neighbor atoms.}
Average sublimation energies of each component in the HEA have a similar sequence {compared to the elemental metals}, however, the discrepancy of bond strength among components is mitigated. 
{The standard deviation of sublimation energies of the elemental metals is 0.50 eV, while that of the HEA atoms is only 0.28 eV.}
The trend of homogenization of bonds is also  reflected in the critical voltage. As shown in Tab.~\ref{tab:hea_v_tab}, the critical voltage of the HEA equals the average of the critical voltages of elemental metals of the components following the rule of mixture, which is an important result that we will revisit {in Sec.~\ref{sec:iap}} in the case of Cu-Ni alloy, where we discuss the assessment of the reliability of interatomic potentials.

{Although a homogenization effect of bonds is observed in the HEA, the copper atoms still have the weakest bonds among the alloy components, and their average sublimation energy has the largest deviation $-8.3$\% from the average of all atoms in the sample $-4.95$ eV. In contrast, the deviations of Ni, Co, Fe and Cr are 4.8\%, 2.6\%, 0.59\% and 0.42\% respectively. The much weaker bonding of copper atoms makes them the easiest ones to be evaporated in the HEA, and their preferential evaporation causes a rough tip surface with dimples and evaporation of clusters (separation at the ``weakest link'', (see the inset in Fig.~\ref{fig:sim_tsong}b). With our carefully controlled voltage-driven evaporation, we never observe cluster evaporation in elemental metals. This finding explains and is in agreement with experimental observations on a similar, compositionally homogeneous equiatomic FeNiCrCoMn HEA, where in voltage mode an unusually high fraction of 30\%-40\% multiple vs.\ single hits is observed \cite{MUNIANDY2021}, which is more than an order of magnitude larger than what is typically observed for elemental metal tips \cite{gault2012fraction}.}

{As a final note, we do not observe any recognizable zone lines or poles in the desorption map of the HEA. This agrees with the general expectation, since for example in highly alloyed steels, zone lines and poles start to wash out and disappear with increasing solute concentration \cite{miller1991local,miller2000thedevelopment,yao2016afiltering}. However, at this point, we cannot make any comparison to experiments so far, as no desorption map of HEA has been reported in the literature. }

\newcolumntype{L}[1]{>{\raggedright\arraybackslash}m{#1}}
\newcolumntype{C}[1]{>{\centering\arraybackslash}m{#1}}
\newcolumntype{R}[1]{>{\raggedleft\arraybackslash}m{#1}}

\begin{table}
\caption{\label{tab:hea_v_tab}
{Comparison of elemental vs.~alloy critical voltages of components of the Fe$_{18.8}$Ni$_{21.9}$Cr$_{18.9}$Co$_{21.8}$Cu$_{18.6}$ high entropy alloy.}
}
\begin{ruledtabular}
\centering
    \begin{tabular}{L{3cm}L{0.8cm}L{0.8cm}L{0.8cm}L{0.8cm}L{0.8cm}}
    Component & Ni & Cr & Co & Fe & Cu\\ 
    \hline
    {Concentration (\%)} & 21.9 & 18.9 & 21.8 & 18.8 & 18.6\\ 
    {Critical voltages of elemental metals (V)} & 342.1 & 317.0 & 311.4 & 284.0 & 242.7\\
    \end{tabular}
    \medskip
    \begin{tabular}{L{3cm}L{0.8cm}L{2.4cm+4\tabcolsep}L{0.8cm}}
    {Weighted average of critical voltages (V)} & 301.2 & {Critical voltage of HEA (V)} & 298.4\\ 
    \end{tabular}
\end{ruledtabular}
\end{table}

\subsection{\label{sec:hit pattern}Field desorption pattern}
\label{section:desorption pattern}
By considering the important role of interatomic interactions during {field evaporation}, we are able to reproduce a key feature of field desorption patterns observed in experiments -- enhanced zone lines. As a demonstration, we make a comparison between Ni desorption patterns obtained by experiment, classical finite element APT simulation, and the present TAPSim-MD approach respectively.

\begin{figure*}[!htb]
    \centering
    \includegraphics[width=0.9\linewidth]{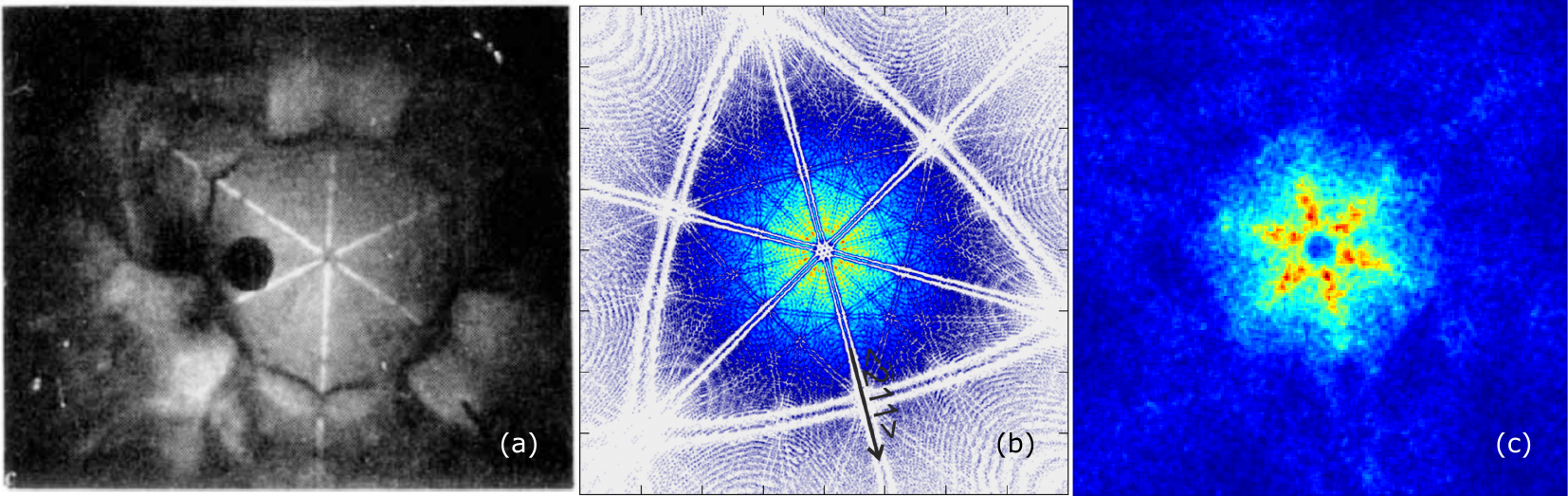}
    \caption{Field desorption patterns obtained for an $\langle$111$\rangle$ oriented Ni tip in (a) experiment, the figure is reprinted from \cite{waugh1975field}; (b) TAPSim simulation for an fcc $\langle$111$\rangle$ oriented structure (tip radius = 25 nm), the figure is reprinted from \cite{oberdorfer2015applications}; (c) proposed TAPSim-MD simulation (tip radius = 8 nm).}
    \label{fig:ni_hit_pattern}
\end{figure*}

In experiments, zone lines with high intensity of hit events can be observed for many metals, such as for the case of a $\langle 111 \rangle$ oriented Ni field desorption pattern shown in Fig.~\ref{fig:ni_hit_pattern}a. This enhanced-zone-line feature of {field desorption} patterns was first reported by Waugh et al.\ about 45 years ago in many materials  \cite{waugh1975field}, and a ``roll-up'' model was then proposed to explain this feature \cite{waugh1976investigations}. However, this conjecture has not been validated yet, and no simulation has been previously able to reproduce this feature. Field desorption patterns simulated by a traditional electrostatic approach \cite{vurpillot1999theshape,geiser2009asystem,oberdorfer2013afullscale,rolland2015ameshless} always display broad and depleted stripes around zone lines. Without considering the local bonding environment of the chemical species, the patterns only depend on the crystal structures of the materials. Figure~\ref{fig:ni_hit_pattern}b is the desorption pattern for the $\langle$111$\rangle$ oriented face centered cubic structure, which is found by the classical TAPSim method to look identical for all fcc-materials \cite{oberdorfer2013afullscale}. In contrast, results obtained here by the TAPSim-MD approach find for the first time zone-line enhancement in agreement with the experiments. Figure \ref{fig:ni_hit_pattern}c is the simulated field desorption pattern of an $\langle$111$\rangle$ oriented Ni tip by TAPSim-MD. The virtual tip is 8 nm in radius. The desorption pattern was obtained by evaporating about 50 layers of atoms. Regions with a warmer color have a higher intensity of atom hit events. {The field desorption pattern} has a three-fold symmetry, and the intensity of atom hit events is higher around the three $\langle$110$\rangle$ zone lines, which agrees with the experimental result in Fig.~\ref{fig:ni_hit_pattern}a. 
{Our simulations demonstrate clearly that the electric field is perpendicular to the envelope of the surface. On the other hand, interatomic forces are anisotropic, and their overall directions are governed by the crystallographic structure. Because of that, the electrostatic desorption force is not always antiparallel to the adhesive interatomic force at the moment of desorption, making the atoms around certain zone lines leave the surface with a non-zero lateral velocity, which accumulates them along certain zone lines. The full quantitative description of the exact mechanism of the enhanced zone lines exceeds the scope of the current publication and will be described in \cite{W-Paper}}.

\section{\label{sec:iap}Influence of interatomic potentials on simulations}

As mentioned before, the selection of the interatomic potential is crucial to the MD simulation. To demonstrate its impact, we compare the results from two common EAM potentials \cite{foiles1985calculation,Onat2013anoptimized} for the Cu-Ni binary alloy system and assess the physical sensibility of their results. 

The virtual tips for these simulations are cylinders with 4 nm in radius and 12 nm in height. Nine virtual tips were generated with randomly arranged Ni impurities with concentrations of 0\%, 5\%, 15\%, 30\%, 50\%, 70\%, 85\%, 95\%, 100\%. Evaporation from these tips was studied with the interatomic potential by Foiles et al.~\cite{foiles1985calculation} in comparison to the potential by Onat et al.~\cite{Onat2013anoptimized}. The important difference between the potentials is that Foiles uses two elemental EAM potentials where the cross term $\phi_{AB}$ of interactions between two species is defined as the geometric average of the elemental pair potentials, whereas Onat fits a dedicated cross term for the alloy potential. 

\begin{figure}[!htb]
    \centering
    \includegraphics[width=\linewidth]{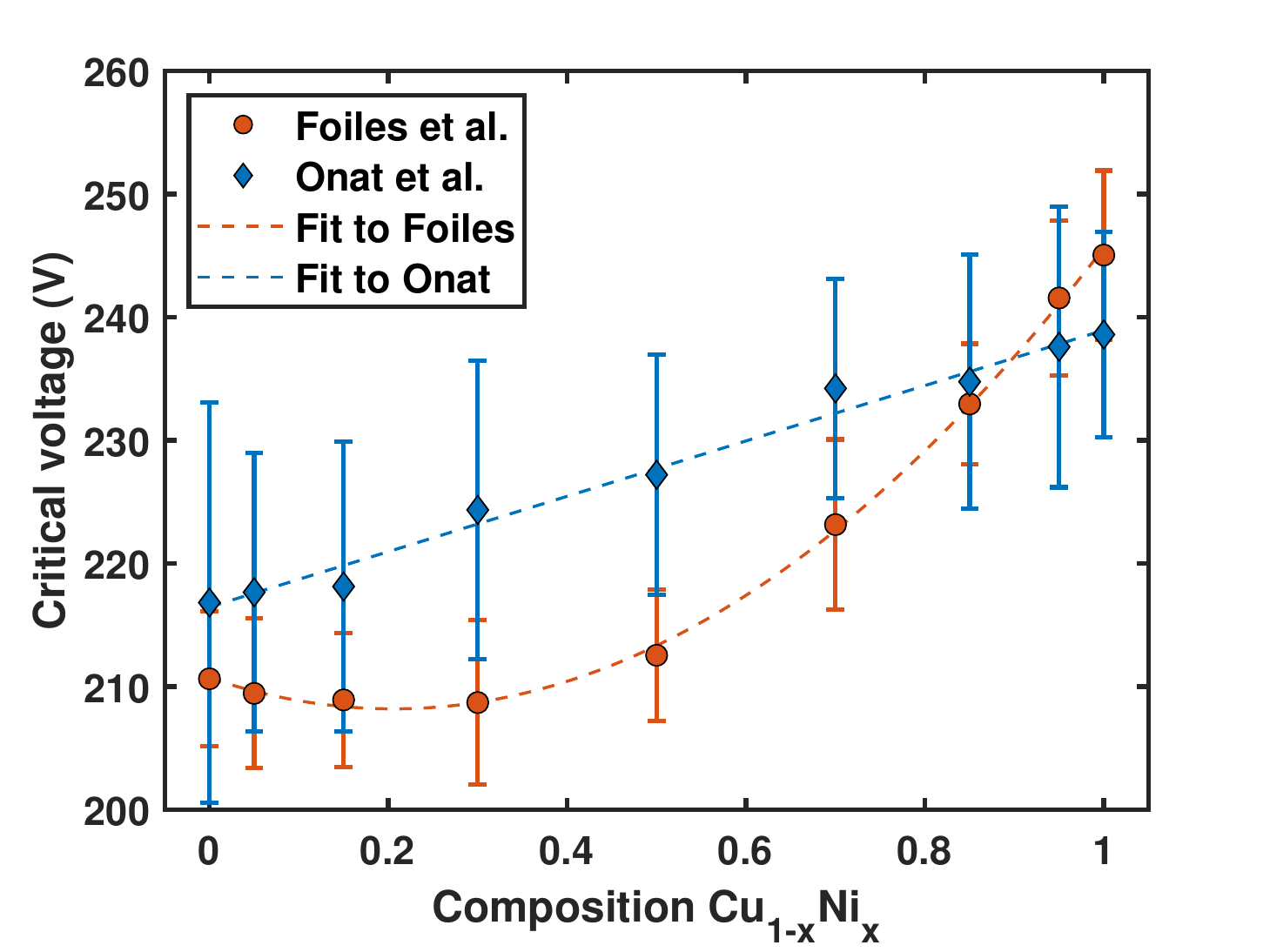}
    \caption{Critical voltages of Cu$_{1-x}$Ni$_x$ alloys obtained by using interatomic potentials from \cite{foiles1985calculation} and \cite{Onat2013anoptimized}. Error bars represent the mean squared errors of the critical voltages.}
    \label{fig:cuni_voltages}
\end{figure}

The critical voltages in the simulations with the potentials were determined in the same way described before. For an isomorphous system such as Cu-Ni, properties follow in general the mixing rule as we have just shown for the case of {an HEA in Sec.~\ref{sec:sublimation energy}}, Fig.~\ref{fig:sim_tsong}b. As shown in Fig.~\ref{fig:cuni_voltages}, the Foiles potential produces critical voltages that have a pronounced quadratic dependence on the composition, while the linear relationship that we expect according to our HEA simulations is reproduced by the Onat potential. Considering the nature of the cross potential as a geometric average in the  potential from Foiles, the quadratic relation should simply originate from the number of bonds in a binary solid solution, which quadratically depends on the composition \cite{oberdorfer2019bond}. However, the potential from Onat shows a linear dependence consistent with general expectations, and our results for the miscible elements in the FeNiCrCoCu high entropy alloy also display a ``mixed'' critical voltage. Compared to the dedicated cross term fitted in Onat's potential, the definition of the cross term as the geometric average in Foiles's is too simplified to accurately describe interactions between Cu and Ni atoms, which is the reason why later alloy approaches in EAM do not use it anymore \cite{Ward-RAMPAGE,Roth-RAMPAGE,Hegde-RAMPAGE,johnson1989alloy}. Since the ZBEF of a species is directly connected to its sublimation energy which in turn is closely related to the environment as was already recognized early in e.g.~the ``M\"uller formula'' \cite{yao2015effects}, a correct description of interatomic interactions is significantly important to obtaining reliable sublimation energies.

\section{\label{sec:compute}Computational performance}

With regard to the computational performance, the combination of a traditional finite element solver with LAMMPS results in a computationally demanding simulation, whose performance also limits the size of the virtual sample. So far, the largest tip used in the simulation is 15 nm in radius with 2,005,877 atoms. The computation speed of the MD part largely depends on the type of the chosen interatomic potentials. We use EAM potentials in most of the cases to balance the efficiency and accuracy of the simulation. 
{The computational effort increases significantly with the number of atoms in the virtual tip using the same interatomic potential on the Owens HPC system at the Ohio Supercomputer Center \cite{OhioSupercomputerCenter1987}, making the use of highly parallel supercomputers indispensable. 
To talk about scaling more specifically, since the number of atoms in the top surface of the tip scales with the square of the tip radius, if we want to evaporate the same number of atom layers for different tip sizes, the simulation time required for the tip with a larger radius will be even longer. As for our examples, in the case of $\langle 110\rangle$ oriented elemental tungsten with the EAM potential from Marinica et al.~\cite{wpotential},
it takes approximately 19.97 CPU hours to evaporate a layer of atoms (604 atoms) in Tip A (28,204 atoms, 4 nm in radius), 331.94 CPU hours to evaporate a layer of atoms (1,110 atoms) in Tip B (103,363 atoms, 5 nm in radius), and 3767.18 CPU hours to evaporate a layer of atoms (4,432 atoms) in Tip C (484,213 atoms, 10 nm in radius), suggesting order-$N$ scaling of CPU time with the number of atoms in a layer, or alternatively, $N^2$ scaling with the tip radius. On our system, the dependence of CPU hours on evaporated number of atoms can be described with $R^2=0.998$ by $t_{\rm CPU} [{\rm CPU-h}]= N - 673$.}

\section{\label{sec:conclusion}Conclusion}

We have introduced TAPSim-MD as a fully physical simulation tool for field desorption. By integrating LAMMPS with the field and surface-charge solver TAPSim, the new TAPSim-MD approach is able to evaporate atoms in a virtual tip by performing classical MD simulations driven by both electric-field-induced forces and interatomic forces. This approach was validated from several aspects. We have first shown that the electric field distribution obtained by solving Poisson's equation compares favorably with results from electron holography measurements \cite{migunov2015model}. Second, we demonstrate that the material-dependent critical voltages and resulting steady-state shapes of virtual tips, as well as critical evaporation fields found in elemental metals and {sublimation energies obtained in} a high entropy alloy are in line with previous predictions \cite{tsong1978field} and general expectations. Finally, field desorption patterns displaying enhanced zone lines are successfully reproduced for the first time, in contrast to all previous forward-simulation work \cite{vurpillot2015modeling}. By eliminating the ad-hoc assumptions adopted by previous approaches concerning evaporation fields and related criteria, our model is able to capture key characteristics of field evaporation in a fully physics-based {\it ``ab-initio''} way, provided a sensible interatomic potential is chosen as illustrated for Cu-Ni alloys. Therefore, sufficient benchmarking and validation needs to be done to balance prediction fidelity and computational efficiency.

The development and validation of TAPSim-MD described in this paper as well as its general nature which is not restricted to a specific material system such as metals enables applications to in principle any arbitrary material as long as a reasonable potential can be identified, as well as examining questions that could not be answered before. Obvious examples include a better understanding of the influence of inhomogeneous bonding on the evaporation sequence and related artifacts as well as field desorption maps. In addition, the full knowledge of the initial atom positions and the resulting detector hit maps enables uncertainty quantification to benchmark classical reconstruction methods.

\section*{Acknowledgements}

This work is sponsored by AFOSR (PM Dr. Ali Sayir) under Award No. FA9550-14-1-0249 and FA9550-19-1-0378. Computing resources are supplied and maintained by the Ohio Supercomputer Center \citep{OhioSupercomputerCenter1987} under Grant No. PAA0010. Results of the simulations were visualized with the programs OVITO \citep{ovito} and ParaView \citep{paraview}. We thank Dr.\ David Seidman for helpful discussions concerning the experimental situation of field evaporation of gold.

\end{document}